\documentclass{article}
\usepackage{amsmath,amsthm,amsfonts,graphicx}
\usepackage{url,calc,multirow,verbatim}

\usepackage[ruled,nothing]{algorithm}
\usepackage{algorithmic}
\newcommand{\GLOBAL}{\item[{\textbf{Global:}}]}

\makeatletter
\newcommand{\IFTHEN}[3][default]{\ALC@it\algorithmicif\ #2\
  \algorithmicthen\ #3\
  \ifthenelse{\boolean{ALC@noend}}{}{\algorithmicendif\ } \ALC@com{#1}}
\makeatother

\newcommand{\linbox}{{\sc LinBox}}
\newcommand{\kaapi}{{\sc Kaapi}}

\newcommand{\Z}{\ensuremath{\mathbb Z}}
\newcommand{\ZnZ}[1]{\leavevmode\kern.1em\raise.0ex
  \hbox{\Z}\kern-.1em
  /\kern-.15em\lower.3ex\hbox{\ensuremath{#1}}\mbox{\Z}}

\urldef\jgdemail\url{Jean-Guillaume.Dumas@imag.fr}
\urldef\tgemail\url{Thierry.Gautier@inrialpes.fr}
\urldef\jlremail\url{Jean-Louis.Roch@imag.fr}

\newtheorem{exmp}{Example}
\title{Generic design of Chinese remaindering schemes}
\author{Jean-Guillaume Dumas\thanks{Laboratoire J. Kuntzmann, Universit\'e de
  Grenoble. 51, rue des Math\'ematiques, umr CNRS 5224, bp 53X, F38041
  Grenoble, France, \jgdemail. Part of this work was done while the first author was visiting the Claude Shannon Institute of the University College Dublin, Ireland, under a CNRS grant.}
\and Thierry Gautier\thanks{Laboratoire LIG, Universit\'e de
  Grenoble. umr CNRS, F38330 Montbonnot, France. \tgemail, \jlremail. Part of this work was done while the second author was visiting the ArTeCS group of the  University Complutense, Madrid, Spain.}
\and Jean-Louis Roch\footnotemark[2]}

\begin{document}
\date{}

\maketitle
\begin{abstract}
We propose a generic design for Chinese remainder algorithms. 
A Chinese remainder computation consists in reconstructing an
integer value from its residues modulo non coprime integers.
We also propose an efficient linear data structure, a radix ladder, for the
intermediate storage and computations.
Our design is structured into three main modules: a black box residue
computation in charge of computing each residue; a Chinese
remaindering controller in charge of launching the computation and of
the termination decision; an integer builder in charge of the
reconstruction computation.
We then show that this design enables many different forms of Chinese
remaindering (e.g. deterministic, early terminated, distributed,
etc.), easy comparisons between these forms and e.g. user-transparent
parallelism at different parallel grains.
\end{abstract}


\section{Introduction}
Modular methods are largely used in computer algebra to reduce the
cost of coefficient growth of the integer, rational or polynomial
coefficients.
Then Chinese remaindering (or interpolation) can be used to recover
the large coefficient from their modular evaluations by reconstructing an
integer value from its residues modulo non coprime integers.

\linbox\footnote{\url{http://linalg.org}}\cite{jgd:2002:icms} is an
exact linear algebra
library providing some of the most efficient methods for linear
systems over arbitrary precision integers.
For instance, to compute the determinant of a large dense matrix over
the integers one can use linear algebra over word size finite fields
\cite{jgd:2008:toms} and then use a combination of system solving and
Chinese remaindering to lift the result \cite{jgd:2006:det}.  
The Frobenius normal form of a matrix is used to test
two matrices for similarity. Although the Frobenius normal form
contains more in formation on the matrix than the characteristic
polynomial, most efficient algorithms to compute it are based on
computations of characteristic polynomial 
(see for example \cite{Pernet:2007:charp}).
Now the Smith normal form of an integer matrix is useful e.g. in the
computation of homology groups and its computation can be done via the
integer minimal polynomial \cite{jgd:2001:JSC}.
In both cases, the polynomials are computed first modulo several prime
numbers and then only reconstructed via Chinese
remaindering using precise bounds on the integer coefficients of the integer
characteristic or minimal polynomials \cite{jgd:2007:jipam}. 

An alternative to the deterministic remaindering is to terminate the
reconstruction early when the actual integer result is smaller than
the estimated bound
\cite{Emiris:1998:CIC,jgd:2001:JSC,Kaltofen:2002:OSV}.
There after the reconstruction stabilizes for some modular iterations,
the computation is stopped and gives the correct answer with high
probability.

In this paper we propose first in section~\ref{sec:radladder} a linear
space data structure enabling fast computation of Chinese
reconstruction. Then we propose in section~\ref{sec:cra} to structure
the design of a generic pattern of Chinese remaindering into three
main modules: 
a black box residue
computation in charge of computing each residue; a Chinese
remaindering controller in charge of launching the computation and of
the termination decision; an integer builder in charge of the
reconstruction computation.
We show in section \ref{sec:ET} that this design enables many
different forms of Chinese
remaindering (e.g. deterministic, early terminated, distributed,
etc.) and easy comparisons between these forms. We
show then in section~\ref{sec:parallel} that this structure provides
also an easy and efficient way to provide user-transparent parallelism
at different parallel grains. 
Any parallel paradigm can be implemented provided that it fulfills
the defined controller interface.
We here chose to use
\kaapi\footnote{\url{http://kaapi.gforge.inria.fr}}\cite{Gautier:2007:kaapi}
to show the efficiency of our approach on distributed/shared
architectures. 

\section{Radix ladder: linear structure for fast Chinese
  remaindering}\label{sec:radladder}
\subsection{Generic reconstruction}
We are given a black box function which computes the 
evaluation of an integer $R$ modulo any number $m$ (often a prime number). 

To reconstruct $R$, 
we must have enough evaluations $r_j \equiv R \mod m_j$ modulo
coprimes $m_j$. 
To perform this reconstruction, we need two by two liftings with
$U \equiv R \mod M$ and $V \equiv R \mod N$ as follows:
\begin{equation}\label{eq:atomic}
R_{MN} = U + (V-U)\times(M^{-1} \mod N) \times M.
\end{equation}

We will need this combination most frequently in two different settings:
when $M$ and $N$ have the same size, and when $N$ is of size $1$.
The first generic aspect of our development is that for both cases,
the same implementation can be fast.

We first need a complexity model. We do not give much details
on fast integer arithmetic in this paper, instead our point is to show the
genericity of our approach and that it facilitates experiments in
order to obtain goods practical efficiency with any underlying
arithmetic. 
Therefore we propose to use a very
simplified model of complexity where division/inverse/modulo are
slightly slower than multiplication. We denote by $m_\alpha l^\alpha$
the complexity of integer
multiplication of size $l$ with $1 < \alpha \leq 2$, and ranging from
$2l^2$ for classical
multiplication to $O(l^{1+\epsilon})$ for FFT-like algorithms and by
$d_\alpha l^\alpha$ the complexity of division, ranging also from
$3l^2$ to $O(l^{1+\epsilon})$.
We refer to e.g. the GMP
manual\footnote{\url{http://gmplib.org/gmp-man-4.3.0.pdf}} or 
\cite{Granlund:1994:divis,VonzurGathen:1999:MCA} for more accurate
estimates. 

\begin{table}[ht]\center
\begin{tabular}{|c|c|c|c|}
\hline
Size of operands & Mul. & Div. & CRT \\
\hline
$l \times 1$ & $l$ & $3l$ & $8l+O(1)$ \\
$l \times l$ & $m_\alpha l^\alpha$ & $d_\alpha l^{\alpha}$ & $2(m_\alpha+d_\alpha)l^\alpha+O(l)$\\
\hline
\end{tabular}
\caption{Integer arithmetic complexity model}\label{tab:model}
\end{table}

With this in mind we compute formula (\ref{eq:atomic}) with one
multiplication modulo as follows:
\newcommand{\crtrec}{{\sc Reconstruct}}
\begin{algorithm}[ht]
\caption{\crtrec}\label{alg:atomic}
\begin{algorithmic}[1]
\REQUIRE $U \equiv R \mod M$ and $V \equiv R \mod N$.
\ENSURE $R_{MN} \equiv R \mod M \times N$.
\STATE $U_N \equiv V-U \mod N$;
\STATE $M_N \equiv M^{-1} \mod N$;
\STATE $U_N \equiv U_N\times M_N \mod N$;
\STATE $R_{MN} = U + U_N\times M$;
\IFTHEN{$R_{MN}>M\times N$}{$R_{MN} = R_{MN}-M\times N$}
\end{algorithmic}
\end{algorithm}

Now, if the formula (\ref{eq:atomic}) is computed via algorithm
\ref{alg:atomic} and the operation counts is done using column ``Mul.''
for multiplication and ``Div.'' for division/inverse/modulo, then we
have the complexities given in column "CRT" 
of table~\ref{tab:model}.

\subsection{Radix ladder}
\begin{figure}[H]\center
\includegraphics[width=\columnwidth]{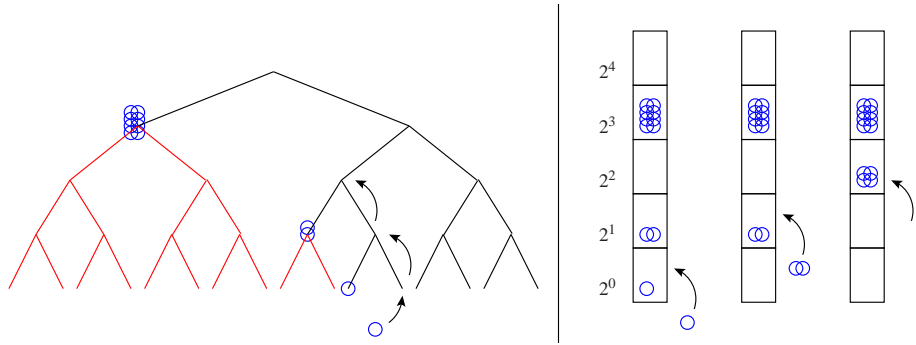}
\caption{A residue going up the radix ladder}\label{fig:radladder}
\end{figure}

Fast algorithms for Chinese remaindering rely on reconstructing pairs
of residues of the same size. A usual way of implementing this is via
a binary tree structure (see e.g. figure~\ref{fig:radladder} left).
But Chinese remaindering is usually an iterative procedure and
residues are added one after the other. Therefore it is possible to
start combining them two by two before the end of the iterations. 
Furthermore, when a combination has been made it contains all the
information of its leaves. Thus it is sufficient to store only the
partially recombined parts and cut its descending branches.
We propose to use a {\em radix ladder} for that task. A radix ladder is a
ladder composed of successive shelves. A shelf is either empty or contains a
modulo and an associated residue, denoted respectively $M_i$ and
$U_i$ at level $i$. Moreover, at level $I$, are stored only residues or
moduli of {\em size} $2^i$. 
New pairs of residue and modulo can be inserted anywhere in the
ladder. If the shelf corresponding to its size is empty, then the pair
is just stored there, otherwise it is combined with occupant of the
shelf, the latter is dismissed and the new combination tries to go one
level up as shown on algorithm~\ref{alg:radladder}.

\newcommand{\radladder}{{\sc RadixLadder}}
\begin{algorithm}[ht]
\caption{\radladder.{\bf insert$(U,M)$}}\label{alg:radladder}
\begin{algorithmic}[1]
\REQUIRE $U \equiv R \mod M$ and a Radix ladder
\ENSURE Insertion of $U$ and $M$ in the ladder., 
\FOR{$i=size(M)$ {\bf while~} Shelf$[i]$ is not empty}
	\STATE $U,M:=$\crtrec$( U \mod M, U_i \mod M_i)$;
        \STATE Pop Shelf$[i]$;
        \STATE Increment $i$;
\ENDFOR
\STATE Push $U, M$ in Shelf$[i]$;
\end{algorithmic}
\end{algorithm}

Then if the new level is empty the combination is stored there,
otherwise it is combined and goes up ... An example of this procedure
is given on figure~\ref{fig:radladder}.

Then to recover the whole reconstructed number it is sufficient to
iterate through the ladder from the ground level and make all the
encountered partial results go to up one level after the other to the
top of the ladder. As we will see in section~\ref{ssec:build},
\linbox-1.1.7 contains
such a data structure, in
{\small \url{linbox/algorithms/cra-full-multip.h}}.

An advantage of this structure is that it enables insertion of any
size pair with fast arithmetic complexity. Moreover, merge of two
ladders is straightforward and we will make an extensive use of that
fact in a parallel setting in
section~\ref{sec:parallel}. 

\begin{algorithm}[ht]
\caption{\radladder.{\bf merge}}\label{alg:radmerge}
\begin{algorithmic}[1]
\REQUIRE Two radix ladders $RL_1$ and $RL_2$.
\ENSURE In place merge of $RL_1$ and $RL_2$.
\FOR{$i=0$~{\bf to}~size$(RL_2)$}
\STATE $RL_1$.insert($RL_2$.Shelf$[i]$);
\ENDFOR
\STATE Return $RL_1$
\end{algorithmic}
\end{algorithm}

\section{A Chinese remaindering design 
pattern}\label{sec:cra}
The generic design we propose here comes from the observation that
there are in general two ways of computing a reconstruction: a
deterministic way computing all the residues until the product of
moduli reaches a bound on the size of the result ; or a probabilistic
way using early termination. We thus propose an abstraction of the
reconstruction process in three layers: a black box function produces
residues modulo small moduli, an integer builder produces
reconstructions using algorithm~\ref{alg:radladder}, and a Chinese
remaindering controller commands them both. 

Here our point is that the controller is completely generic where the
builder may use e.g. the radix ladder data structure proposed in
section~\ref{sec:radladder} and has to implement the termination
strategy. 

\subsection{Black box residue computation}
In general this consists in mapping the problem from $\Z$ to $\ZnZ{m}$
and computing the result modulo $m$. Such black boxes are defined
e.g. for the determinant, valence, minpoly, charpoly, linear system
solve as function objects \url{IntegerModular*} (where \url{*} is one of
the latter functions) in the {\small \url{linbox/solutions}} 
directory of \linbox-1.1.7.

\subsection{Chinese remaindering controller}
The pattern we propose here is generic with respect to the termination
strategy and the integer reconstruction scheme. 
The controller must be able to initialize the data structure via the
builder ; generate some coprime moduli ; apply the black box function
; update the data structure ; test for termination and output the
reconstructed element. The generations of moduli and the black box are
parameters and the other functionalities are provided by any builder. 
Then the control is a simple loop. Algorithm \ref{alg:control} shows
this loop which contains also the whole interface of the Builder.
\begin{algorithm}[ht]
\caption{{\sc CRA-Control}}\label{alg:control}
\begin{algorithmic}[1]
\STATE $Builder$.{\bf initialize}$()$;
\WHILE{$Builder$.{\bf notTerminated}$()$}
\STATE $p:=Builder.${\bf nextCoPrime}$()$;	
\STATE $v:=BlackBox$.{\bf apply}$(p)$;
\STATE $Builder$.{\bf update}$(v,p)$;
\ENDWHILE
\STATE Return $Builder$.{\bf reconstruct}$()$;
\end{algorithmic}
\end{algorithm}

\linbox-1.1.7 gives an implementation of such a controller,
parametrized by a builder and a black box function as the class 
\url{ChineseRemainder} in 
{\small \url{linbox/algorithms/cra-domain.h}}.

The interface of a controller is to be a function class.
It contains a constructor with a
builder as argument and the functional operator taking as argument a BlackBox,
computing e.g. a determinant modulo $m$, and a moduli generator and
returning an integer reconstructed from the modular computations.
Algorithm \ref{alg:cra} shows the specifications of the \linbox-1.1.7
controller.
\begin{algorithm}[ht]
\caption{C++ {\bf ChineseRemainder} class}\label{alg:cra}
\begin{verbatim}
template<class Builder> struct ChineseRemainder {
ChineseRemainder(const Builder& b) : builder_(b) {}
template<class Function> Integer& operator() (
    Integer	       & res, 
    const Function	& BlackBox) {
        // CRA-Control ...
}           
const Builder& getBuilder() { return builder_; }
protected: Builder builder_;
};
\end{verbatim}
\end{algorithm}
Then any higher-level algorithm will just chose  its builder and its
controller and pass them the modular BlackBox iteration it wants to
lift over the integers.

\subsection{Integer builders}\label{ssec:build}
The role of the builder is to implement the interface defined by
algorithm~\ref{alg:control}.

There are already three of these implementations in \linbox-1.1.7: an
early terminated for a single residue, an early terminated for a vector
of residues and a deterministic for a vector of residues 
(resp. the files {\small \url{cra-early-single.h}},
{\small \url{cra-early-} \url{multip.h}} and 
{\small \url{cra-full-multip.h}} in the
{\small \url{linbox/algorithms}} directory).
Up to now the radix ladder is not a separated class as only this
data structure is currently used and as it is simple enough to inherit
from one of the latter and modify the behavior of the methods.

\begin{figure}[ht]\center
\includegraphics[width=\columnwidth*3/4]{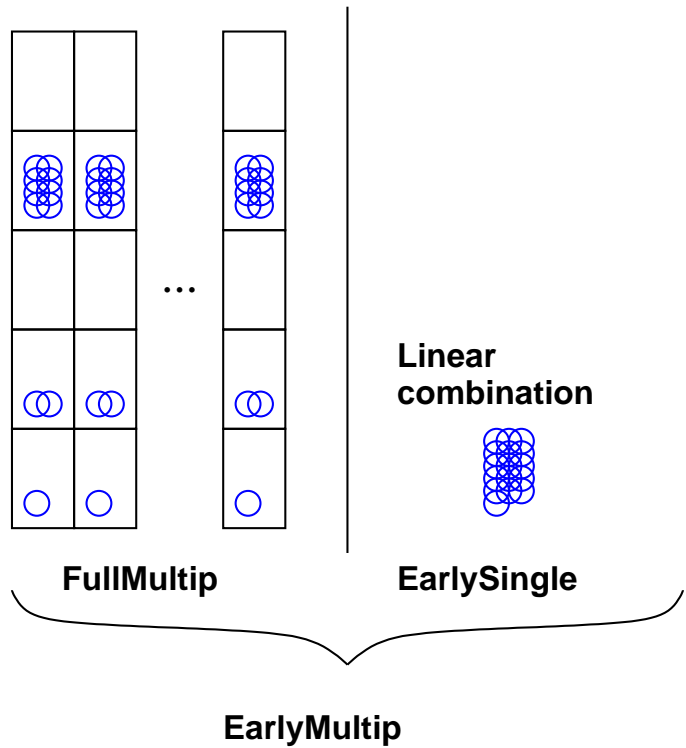}
\caption{Early termination of a vector of residues via a linear combination}\label{fig:earlymultip}
\end{figure}

Actually \url{EarlyMultipCRA} inherits from both
\url{EarlySingleCRA} and \url{FullMultipCRA} as it uses the radix
ladder of \url{FullMultipCRA} for its reconstruction and the early
termination of \url{EarlySingleCRA} to test a linear combination a
the residues to be reconstructed as shown on figure~\ref{fig:earlymultip}

We give more implementation details on the early termination
strategies in sections~\ref{sec:ET} and~\ref{sec:parallel}.

\section{Termination strategies}\label{sec:ET}
Let us sketch here several early termination strategies and show that
our design enables to modify this strategy and only that while the
rest of the implementation is unchanged.
\subsection{Earliest termination}
In a sequential mode, depending on the actual speed of
the different routines of table~\ref{tab:model} 
on a specific
architecture or if the cost of $BlackBox$.{\bf apply} is largely dominant, one can choose to test for termination after each call
to the black box. A way to implement the probabilistic test of \cite[Lemma
3.1]{jgd:2001:JSC} and to reuse every black box apply is to use random
primes as the moduli generator. Indeed then the probabilistic check
can be made with the incoming black box residue computed modulo a
random prime. The reconstruction algorithm of section~\ref{sec:cra} is
then only slightly modified as shown in algorithm \ref{alg:etupd}
\begin{algorithm}[ht]
\caption{{\sc EarlySingleCRA}.{\bf update$(v,p)$}}\label{alg:etupd}
\begin{algorithmic}[1]
\GLOBAL $U \equiv R \mod M$.
\GLOBAL A variable $Stabilization$ initially set to $0$.
\REQUIRE $v \equiv R \mod p$.
\ENSURE $R_{MN} \equiv R \mod M \times p$.
\STATE $u \equiv U \mod p$;
\IF{$u == v$}
\STATE Increment $Stabilization$;
\STATE Return $(U,M \times p)$;
\ELSE
\STATE $Stabilization=0$;
\STATE Return \crtrec$( U \mod M, v \mod p)$;
\ENDIF
\end{algorithmic}
\end{algorithm}
and the termination test becomes simply algorithm \ref{alg:etterm}.
\begin{algorithm}[ht]
\caption{{\sc EarlySingleCRA}.{\bf notTerminated$()$}}\label{alg:etterm}
\begin{algorithmic}[1]
\STATE Return $Stabilization < EarlyTerminationThreshold$;
\end{algorithmic}
\end{algorithm}

In the latter algorithm, $EarlyTerminationThreshold$ is the number of
successive
stabilizations required to get a probabilistic estimate of
failures. It will be denoted $ET$ for the rest of the paper.
This is the strategy implemented in \linbox-1.1.7 in
{\small \url{linbox/algorithms/cra-early-single.h}}.
With the estimates of
table~\ref{tab:model}, the cost of the whole 
reconstruction of algorithm~\ref{alg:control} thus becomes 
\begin{multline}\small
\sum_{i=1}^{t} \left(\text{\bf apply}+8i+O(1) \right) =\\
(t+ET)\text{\bf apply}+4(t+ET)^2+O(t)
\end{multline}
where $t=\lceil \log_{2^\beta}(R)\rceil$ and
$\beta$ is the word size.

This strategy enables the least possible number of calls to
$BlackBox$.{\bf apply}. It it thus useful when the latter dominates the
cost of the reconstruction.

\subsection{Balanced termination} 
Another classic case is when one wants to use fast integer arithmetic
for the reconstruction.
Then the balanced computations are mandatory and the radix ladder
becomes handy.
The problem now becomes the early termination. There a simple strategy
could be to test for termination only when the number of computed
residues is a power of two. In that case the reconstruction is
guaranteed to be balanced and fast Chinese remaindering is also
guaranteed.
Moreover random moduli are not any more necessary for all the
residues, only those testing for early termination need be randomly
generated. This induces another saving if one fixes the other primes
and precomputes all the factors $M_i \times (M_i^{-1} \mod M_{i+1})$. 
There the cost of the reconstruction drops by a factor of $2$ from
$2(m_\alpha+d_\alpha)l^\alpha$ to $(m_\alpha+d_\alpha)l^\alpha$.

The drawback is an extension of the number of black box
applications from $\lceil \log_{2^\beta}(R)\rceil+ET$ to 
the largest power of two immediately superior and thus up to a factor
of $2$ in the number of black box applies.

For the $Builder$, the update becomes just a push in the ladder as
shown on algorithm~\ref{alg:balupd}.
\begin{algorithm}[ht]
\caption{{\sc EarlyBalancedCRA}.{\bf update$(v,p)$}}\label{alg:balupd}
\begin{algorithmic}[1]
\STATE \radladder.{\bf insert}$(v, p)$;
\end{algorithmic}
\end{algorithm}

The termination condition, on the contrary tests only when the number
of residues is power of two as shown on algorithm~\ref{alg:balterm}.
\begin{algorithm}[ht]
\caption{{\sc EarlyBalancedCRA}.{\bf notTerminated$()$}}\label{alg:balterm}
\begin{algorithmic}[1]
\IF{Only one Shelf, Shelf$[i]$, is full}
\STATE Set $U_i$ to Shelf$[i]$ residue;
\FOR{$j=1$  {\bf to~} $EarlyTerminationThreshold$}
	\STATE $p:=${\sc PrimeGenerator}$()$;
        \IF{$(U_i \mod p)~~!=~BlackBox$.{\bf apply}$(p)$}
        	\STATE Return $false$;
        \ENDIF
\ENDFOR
\STATE Return $true$;
\ELSE
\STATE Return $false$;
\ENDIF
\end{algorithmic}
\end{algorithm}

Then, the whole reconstruction of algorithm~\ref{alg:control} now requires:
\begin{equation}\begin{split}
ET \cdot (\text{\bf apply}+3\cdot 2^k)+\sum_{i=0}^{k-1}
\frac{2^k}{2^{i+1}}\left(\text{\bf
  apply}
+(m_\alpha+d_\alpha)2^{i\alpha}\right)\\+(\text{\bf apply}+3\cdot 2^i) = \\
(2^k+k+ET-1)\cdot \text{\bf apply}+\left(2^{k}\right)^\alpha\frac{m_\alpha+d_\alpha}{2^\alpha-2}+O(2^k)
\end{split}
\end{equation}
operations, where now $k=\lceil \log_2(\log_{2^\beta}(R)) \rceil$.

Despite the augmentation in the number of black box applications, 
the latter can be useful, in particular when multiple values are to be
reconstructed.
\begin{exmp}\label{exmp:gauss}
  Consider the Gau{\ss}ian elimination of an integer
  matrix where all the
  matrix entries are larger than $n$ and bounded in absolute value by
  $A_\infty$. We denote $\log_{2^\beta}(A_\infty)$ by $a_\infty$. Suppose one
  would like to compute the rational coefficients of the triangular
  decomposition only by Chinese remaindering (there exist better output
  dependant algorithms, see e.g. \cite{Pernet:2009:hnf}, but the
  latter has the same worst-case complexity).
  Now, Hadamard bound gives
  that the resulting numerators and denominators of the coefficients
  are bounded by
  $\sqrt{nA_\infty^2}^n$. Then the complexity of the earliest strategy
  would be dominated by the reconstruction where the balanced strategy or
  the hybrid strategy of figure~\ref{fig:earlymultip} could
  benefit from fast algorithms:
\begin{table}[H]\center
\begin{tabular}{|l|l|}
\hline
\url{EarlySingleCRA} & $O(n^4a_\infty^2)$\\
\url{EarlyMultipCRA} & $O(n^{\omega+1}a_\infty+n^{2+\alpha}a_\infty^\alpha+n^2a_\infty^2)$\\
\url{EarlyBalancedCRA} & $O(2n^{\omega+1}a_\infty+n^{2+\alpha}a_\infty^\alpha)$\\
\hline
\end{tabular}
\caption{Early termination strategies complexities for Chinese
  remaindered Gau{\ss}ian elimination with rationals}
\end{table}
In the case of small matrices with large entries the reconstruction
dominates and then a balanced strategy is preferable. Now if both
complexities are comparable it might be useful to reduce the factor of
$2$ overhead in the black box applications. This can be done via amortized
techniques, as shown next.
\end{exmp}

\subsection{Amortized termination}
A possibility is to use the $\rho$-amortized control of
\cite{Beaumont:2004:PMAA}: instead of testing for termination at steps
$2^1$, $2^2$, $\ldots$, $2^i$, $\ldots$ the tests are performed at steps
$\rho^{g(1)}$, $\rho^{g(2)}$, $\ldots$, $\rho^{g(i)}$, $\ldots$ with
$1<\rho<2$ and $f$ satisfies $\forall i$, $g(i)\leq i$. 
If the complexity of the modular problem is $C$ and the number of
iterations to get the output is $b$,
\cite{Beaumont:2004:PMAA} give choices for $\rho$ and $g$
which enable to get the result with only $b+\frac{f(b)}{b}$ iterations
and extra $O(f(b))$ termination tests where $f(b)=log_\rho(b)$.

In example~\ref{exmp:gauss} the complexity of the modular problem is
$n^\omega$, the size of the output and the number of iterations is
$na_\infty$ so that strategy would reduce the iteration complexity
from $2n^{\omega+1}a_\infty$ to
$(na_\infty+o(na_\infty))n^{\omega}$ and the overall complexity would
then become:
\begin{center}
\begin{tabular}{|l|l|}
\hline
\multirow{2}{*}{\url{EarlyAmortizedCRA}} &
$O(n^{\omega+1}a_\infty+n^{2+\alpha}a_\infty^\alpha$\\
& \multicolumn{1}{r|}{$+\log(na_\infty)n^{\alpha}a_\infty^\alpha)$}\\
\hline
\end{tabular}
\end{center}

Indeed, we suppose that the amortized technique is used only on a linear
combination, and that the whole
matrix is reconstructed with a \url{FullMultipCRA}, as in figure
\ref{fig:earlymultip}. Then the linear combination has size
$2\log(n)+n\cdot a_\infty$ which is still $O(n\cdot a_\infty)$.
Nonetheless, there is an overhead of a factor $\log(na_\infty)$ in the
linear combination reconstruction since there
might be up to $O(\log(na_\infty))$ values $\rho^{g(i)}$,
$\rho^{g(i+1)}$, $\ldots$ between any two powers of two. Overall this
gives the above estimate.
Now one could use other $g$ functions as long as eq.~\ref{eq:amort} is
satisfied.
\begin{equation}\label{eq:amort}
\begin{cases}
 \left(\rho^{g(i+1)}-\rho^{g(i)}\right) = o(\rho^{g(i)})\\
 \left(\rho^{g(i+k(i))}-\rho^{g(i)}\right) \sim 2^{\lceil \log_2(\rho^{g(i)})
   \rceil},& k(i)=o(\rho^{g(i)})
\end{cases}
\end{equation}

\section{Parallelization}\label{sec:parallel}
All parallel versions of these sequential algorithms have to consider
the parallel merge of radix ladders and 
the parallelization of the loop of the CRA-control
algorithm~\ref{alg:control}. 
Many parallel libraries can be used, namely OpenMP or Cilk
would be good candidates for the parallelization of the embarrassingly
parallel \url{FullMultipCRA}. 
Now in the early termination setting, the main difficulty comes from
the distribution of the termination test. Indeed, the latter depends
on data computed during the iterations. 
To handle this issue we propose an adaptive parallel
algorithm~\cite{TC2006-aha,Europar2008-TRMGB} and use the
\kaapi~library~\cite{DGGRR-pasco07,Gautier:2007:kaapi}. Its
expressiveness in an adaptive setting guided our choice, together with
the possibility to work on heterogenous networks.

\subsection{\kaapi~overview}
\kaapi~is a task based model for parallel computing. It was targeted
for distributed and shared memory computers. The scheduling algorithm
uses
work-stealing~\cite{blumofe96cilk,Arora01threadscheduling,ChaseL05,inproceedingsgautier.grw_fgdidl_07}:
an idle processor tries to steal work to a randomly selected victim
processor. 

The sequential execution of a \kaapi~program consists in pushing and
popping tasks to dequeue the current running processor. Tasks should
declare the way they access the memory, in order to compute, at
runtime, the data flow dependencies and the ready tasks (when all
their input values are produced). During a parallel execution, a ready
task, in the queue but not executed, may be entirely theft and
executed on an other processor (possibly after being communicated
through the network). These tasks are called \textit{dfg tasks} and
their schedule by work-stealing is described
in~\cite{Gautier:2007:kaapi,inproceedingsgautier.grw_fgdidl_07}. 

A task being executed by a processor may be \textit{only partially}
theft if it interacts with the scheduler, in order to e.g. decide which
part of the work is to be given to the thieves. 
Such tasks are called \textit{adaptive tasks} and allows fine grain
loop parallelism.

To program an adaptive algorithm with Kaapi, the programmer has to
specify some points in the code (using \url{kaapi_stealpoint}) or
sections of the code (\url{kaapi_stealbegin}, \url{kaapi_stealend})
where thieves may steal work. 
To guarantee that parallel computation is completed, 
the programmer has to wait for the finalization of the parallel
execution (using \url{kaapi_steal_finalize}). 
Moreover, in order to better balance the work load, the programmer may
also decide to preempt the thieves (send an event via
\url{kaapi_preempt_next}).

\subsection{Parallel earliest termination}
Algorithm~\ref{alg:parcontrol} lets thieves steal any sequence of
primes.
\begin{algorithm}[ht]
\caption{{\sc ParallelCRA-Control}}\label{alg:parcontrol}
\begin{algorithmic}[1]
\STATE $Builder$.{\bf initialize}$()$;
\WHILE{$Builder$.{\bf notTerminated}$()$}
\STATE $p:=Builder.${\bf nextCoPrime}$()$;	\label{lin:genp}
\STATE\label{lin:beg} \textbf{kaapi\_stealbegin}$(\ splitter,\ Builder )$;
\STATE $v:=BlackBox$.{\bf apply}$(p)$;\label{lin:apply}
\STATE $Builder$.{\bf update}$(v,p)$;\label{lin:update}
\STATE {\textbf{kaapi\_finalize\_steal}$()$;}
\STATE\label{lin:end} \textbf{kaapi\_stealend()};
\IF{require synchronization step}\label{lin:sync}
\WHILE{\textbf{kaapi\_nomore\_thief}$()$}\label{lin:nomore}  
\STATE $(list\ of\ v, list\ of\ p):=${\textbf{kaapi\_preempt\_next()};}\label{lin:dopreempt}
\STATE\label{lin:upd} $Builder$.{\bf update}$(list\ of\ v,list\ of\ p)$;
\ENDWHILE
\ENDIF
\ENDWHILE
\STATE Return $Builder$.{\bf reconstruct}$()$;
\end{algorithmic}
\end{algorithm}\\
At line \ref{lin:beg}, the code allows the scheduler to trigger the
processing of
steal requests by calling the $splitter$ function. 
The parameters of \url{kaapi_stealbegin} are the $splitter$ function
and some arguments to be given to its call. 
These arguments\footnote{in or out} can
e.g. specify the state of the computation to modify (here the builder
object plays this role). 
Then, on the one hand, concurrent modifications of the state of
computation by thieves, must be taken care of 
during the control flow between lines \ref{lin:beg} and \ref{lin:end}:
here the computation of the residue could be evaluated by multiple
threads without critical section\footnote{This depends on the
  implementation, most of the LinBox library functions are reentrant}.
On the other hand, after line \ref{lin:end}, 
the scheduler guarantees that no concurrent thief can modify the
computational state when they steal some work. 
Remark that both branches of
the conditional \url{if} at line~\ref{lin:sync} must be executed 
without concurrency: the iteration of the list of thieves 
or the generation of the next random modulo are not reentrant.

The role of the $splitter$ function is to distribute the work among
the thieves.
In algorithm~\ref{alg:splitter}, each thief receives a
\url{coPrimeGenerator} object 
and the $entrypoint$ to execute. 
\begin{algorithm}[ht]
\caption{{\sc Splitter}$(Builder, N, requests[])$}\label{alg:splitter}
\begin{algorithmic}[1]
\FOR{$i=0$ \textbf{to} $N-1$}
\STATE \textbf{kaapi\_request\_reply}$(request[i], entrypoint,$ $\hspace*{1cm}Builder.getCoPrimeGenerator()\ )$;
\ENDFOR
\end{algorithmic}
\end{algorithm}

The \url{coPrimeGenerator} depends on the  \url{Builder} type and
allows the thief to generate a sequence of moduli. 
For instance the \url{coPrimeGenerator} for the earliest termination
contains at one point a single modulo $M$ which 
is returned by the next call of \url{nextCoPrime()} by the
\url{Builder}.

The  $splitter$ function knows the number $N$ of thieves that are
trying to steal work to the same victim. Therefore it allows for a better balance of the work
load.
This feature is unique to \kaapi~when compared to other tools having a
work-stealing scheduler.  

\subsection{Synchronization}
Now, the victim periodically tests the global termination
of the computation (line \ref{lin:sync} of
algorithm~\ref{alg:parcontrol}). 
Depending on the chosen termination method (\url{Early*CRA}, etc.), 
the synchronization may occur at every iteration or after a certain
number of iterations.
The choice is made in order to e.g. amortize the cost of this
synchronization or reduce the arithmetic cost of the reconstruction.
Then each thief is preempted (line~\ref{lin:dopreempt}) and the code
recovers its results before  
giving them to the \url{Builder} for future reconstruction (line
\ref{lin:upd}).

The preemption operation is a two way communication between a victim
and a thief: the victim may pass parameters {\em and} get data from
one thief.
Note that the preemption operation assumes cooperation with the thief
code. The latter being responsible for polling incoming events at
specific points (e.g. where the computational state is safe
preemption-wise).

On the one hand, to amortize the cost of this synchronization, more
primes should be given to the thieves. In the same way, the victim code 
works on a list of moduli inside the critical section 
(at line~\ref{lin:genp} returns a list of moduli, and 
at lines~\ref{lin:apply}-\ref{lin:update} the victim iterates over this list by repeatedly calling \url{apply} and \url{update} methods).
On the other hand, to avoid long waits of the victim during preemption, 
each thief should test if it has been preempted to return quickly its results 
(see next section).

\subsection{Thief entrypoint}
Finally, algorithm~\ref{alg:thiefentrypoint} returns both the sequence
of residues and the sequence of primes that where given to the BlackBox. 
This algorithm is very similar to algorithm~\ref{alg:parcontrol}.
\begin{algorithm}[ht]
\caption{{\sc Thief's EntryPoint(M)}}\label{alg:thiefentrypoint}
\begin{algorithmic}[1]
\STATE $Builder$.{\bf initialize}$()$;
\STATE $list\ of\ v$.{\bf clear}$()$;
\STATE $list\ of\ p$.{\bf clear}$()$;
\WHILE{$Builder$.{\bf CoPrimeGenerator}$()$ not empty}
\IFTHEN{{\bf kaapi\_preemptpoint}$()$}{break;}\label{lin:premp}
\STATE $p:=Builder.${\bf nextCoPrime}$()$;	
\STATE\label{lin:tbeg} \textbf{kaapi\_stealbegin}$(\ splitter,\ Builder )$;
\STATE $list\ of\ p$.push\_back$(p)$;
\STATE $list\ of\ v$.push\_back$(BlackBox$.{\bf apply}$(p))$;
\STATE\label{lin:tend} \textbf{kaapi\_stealend()};
\ENDWHILE
\STATE \textbf{kaapi\_stealreturn} $(list\ of\ v,list\ of\ p)$;
\end{algorithmic}
\end{algorithm}

Lines~\ref{lin:tbeg} and \ref{lin:tend} define a section of code that could be concurrent with steal requests.
At line~\ref{lin:premp}, the code tests if a preemption request has
been posted by algorithm~\ref{alg:parcontrol} at line~\ref{lin:dopreempt}.
If this is the case, then the thief aborts any further
computation and the result is only a partial set of the initial work
allocated by the $splitter$ function.


\subsection{Efficiency}
These parallel versions of the Chinese remaindering have been
implemented using \kaapi~transparently from the \linbox~library: one has
just to change the sequential controller {\small \url{cra-domain.h}}
to the parallel one.

In \linbox-1.1.7 some of the sequential algorithms which make use of
some Chinese remaindering are the determinant, the
minimal/characteristic polynomial and the valence, see e.g.
\cite{Kaltofen:2002:OSV,jgd:2001:JSC,jgd:2005:charp,jgd:2007:jipam}
for more details.

We have performed these preliminary experiments on an 8 dual core
machine (Opteron 875, 1MB L2 cache, 2.2Ghz, with 30GBytes of main
memory). 
Each processor is attached to a memory bank and communicates to its
neighbors via an hypertransport network. 
We used g++ 4.3.4 as C++ compiler and the Linux kernel was the 2.6.32 
Debian distribution. 
 
All timings are in seconds. 
In the following, we denote by $T_{seq}$ the time of the sequential
execution and by $T_p$ the time of the parallel execution for $p=8$
or $p=16$ cores. 
All the matrices are from ``Sparse Integer Matrix
Collection'' (SIMC)\footnote{\url{http://ljk.imag.fr/CASYS/SIMC}}. 

Table~\ref{tab:det} gives the performance of the parallel computation
of the determinant for small invertible matrices (less than a second)
and larger ones (an hour CPU) of the \url{SIMC/SPG} and
\url{SIMC/Trefethen} collections.
\begin{table}[ht]
\begin{small}
$$
\hspace*{-3ex}
\begin{array}{|c|c||c|c|c|}
\hline
\bf Matrix & \bf d, r &  \bf T_{seq} [k] & \bf T_{p=8} [k] & \bf T_{p=16} [k] \\
\hline
\hline
ex-1 & 560,\ 8736 & 0.336 [4] & 0.386 [60] & 0.533 [124] \\
ex-3 & 2600,\ 71760 & 914.28 [183] &  247.75 [303] & 102.75 [263] \\
\hline
t-150 & 150,\ 2040 & 0.26 [59] & 0.14 [135] & 0.13 [263] \\
t-300 & 300,\ 4678 & 2.90 [138] & 0.88 [143] & 0.48  [255] \\
t-500 & 500,\ 8478 & 17.65 [249]& 3.05 [319] & 1.57 [255] \\
t-700 & 700,\ 12654 & 58.25 [367]  & 11.45 [386] & 7.56 [376] \\
t-2000 & 2000,\ 41907 & 3208.59 [1274]  & 708.506[1295] & 434.31[1286] \\
\hline
\end{array}
$$
\end{small}
\caption{Timings for the computation of the determinant. $d$ is the
  dimension of the matrix, $r$ the number of non-zero coefficients, $[k]$ is the minimal number of primes observed for the Chinese remaindering using $p$ cores.
}
\label{tab:det}
\end{table}\\
The small instance (ex-1) needed very few primes to reconstruct
integer the solution. There, we can see the overhead of parallelism:
this is due to some extra synchronizations and also to the large
number of unnecessary modular computations before realizing that early
termination was need. Despite this (124 modular computations with 16
cores against 4 modular applications in sequential) the overhead was
less than $0.2$ second.
Now, for matrix ex-3,  the speed up for 16 cores was a little bit more
than $8$: with the large amount of memory, we assume that the memory bandwidth
is the bottleneck when two cores on the same processor share the
memory interconnection.

\begin{table}[ht]\center
\begin{small}
$$
\hspace*{-2ex}
\begin{array}{|c|c||c|c|c|}
\hline
\bf Matrix & \bf d &  \bf T_{seq} [k] & \bf T_{p=4} [k] & \bf T_{p=8} [k] \\
\hline
\hline
ch88.b5 & 564480 \times 376320 & 73.78 [5] & 40.53 [6] & 47.58 [5] \\
\hline
n4c5.b5 & 4340 \times 2852 & 	5.62 [16] & 2.09 [17] & 3.09 [35] \\
n4c5.b6 & 4735 \times 4340 & 25.15 [32] & 10.07 [33] & 9.12 [35] \\
n4c5.b7 & 3635 \times 4735 & 29.87 [38] & 12.17 [42] & 9.07 [40] \\
n4c6.b5 & 51813 \times 20058 & 92.29 [18] & 34.16 [22] & 35.45 [29] \\
\hline
\end{array}
$$
\end{small}
\caption{Times for the computation of the valence. $d$ is the dimension of the matrix, $[k]$ is the minimal number of primes observed for the Chinese remaindering using $p$ cores.}
\label{tab:val}
\end{table}
Table~\ref{tab:val} shows
the performance of the parallel computation of the valence of $A\,^t A$.
Matrices in this table come from the \url{SIMC/Homology} collection and
were used in~\cite{jgd:2001:JSC}, where the
parallelism was ad-hoc. 

Further analysis is required to identify hot spots in the executions
in order to improve the efficiency at this computational grain. 
First at all, we need to increase the size of the critical section between
\url{kaapi\_stealbegin} \url{kaapi\_stealend}: \textit{e.g.} by generating
in parallel random primes. 
We already found some bottlenecks due to memory allocation that may be
removed. 
Besides, it remains difficult on such architecture to
control the computational affinity with the mapping of data in
memory. 

\section{Conclusion}
We have proposed a new data structure, the radix ladder, 
capable of managing several kinds of Chinese reconstructions.

\begin{figure*}[htb]\center
\includegraphics[width=\textwidth]{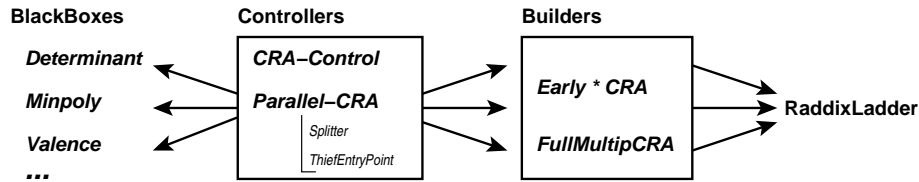}
\caption{Generic Chinese remaindering scheme}\label{fig:cradesign}
\end{figure*}

Then, we have defined a new generic design for Chinese remaindering 
schemes. It is summarized on figure \ref{fig:cradesign}. Its main
feature is the definition of a builder interface in charge of the
reconstruction. This interface is such that any of termination
(deterministic, early terminated, distributed, etc.) can be handled by
a CRA controller. It enables to define and test remaindering
strategies while being transparent to the higher level routines. 
Indeed we show that the Chinese remaindering can just be 
a plug-in in any integer computation.

We also provide in \linbox-1.1.7 an implementation of the ladder,
several implementations for different builders and a
sequential controller. Then we tested the introduction of a parallel
controller, written with \kaapi, without any modification of the
\linbox~library. 
The latter handles the difficult issue of distributed early
termination and shows good performance on a SMP machine.

In parallel, some improvement could be made to the early termination
strategy in particular when the BlackBox is fast compared to the
reconstruction and when balanced and amortized techniques are required.
Also, output sensitive
early termination is very useful for rational reconstruction, see
e.g. \cite{Khodadad:2006:FRF} and thus the latter
should benefit from this kind of design. 

\bibliographystyle{abbrv}
\bibliography{crt} 

\end{document}